\documentclass[aps,prl,twocolumn,superscriptaddress,amsfont,graphicx,nofootinbib,preprintnumbers]{revtex4}%
\usepackage{color,graphicx,epsfig}
\usepackage{ifpdf}
\usepackage{amsmath}
\usepackage{bm}
\usepackage{color}
\usepackage[english]{babel}
\usepackage{graphicx}%
\usepackage{amsfonts}%
\usepackage{amssymb}
\usepackage{braket}
\usepackage{hyperref}
\usepackage{enumerate}

\definecolor{nicered}{rgb}{0.7,0.1,0.1}
\definecolor{nicegreen}{rgb}{0.1,0.5,0.1}
\hypersetup{colorlinks,citecolor= nicegreen,linkcolor= nicered}

\newcommand{\beq}{\begin{equation}}
\newcommand{\eeq}{\end{equation}}
\newcommand{\bea}{\begin{eqnarray}}
\newcommand{\eea}{\end{eqnarray}}

\definecolor{Red}{rgb}{1.,0.,0.}

\def\taun{{\cal T}_N}
\def\tauone{{\cal T}_1}
\def\tauncut{{\cal T}_N^{cut}}
\def\tauonecut{{\cal T}_1^{cut}}

\def\OMIT#1{}

\begin{document}

\def\Peking{Center for High-Energy Physics, Peking University, Beijing, 100871, China}
\def\Maryland{Maryland Center for Fundamental Physics, University of Maryland, College Park, Maryland 20742, USA}
\def\Argonne{High Energy Physics Division, Argonne National Laboratory, Argonne, IL 60439, USA}
\def\Northwestern{Department of Physics \& Astronomy, Northwestern University, Evanston, IL 60208, USA}
\def\Fermilab{Fermilab, P.O.Box 500, Batavia, IL 60510, USA}

\preprint{FERMILAB-PUB-15-210-T}

\title{Higgs boson production in association with a jet at NNLO using jettiness subtraction}

\author{Radja Boughezal}     
\email[Electronic address:]{rboughezal@anl.gov}
\affiliation{\Argonne}

\author{Christfried Focke}
\email[Electronic address:]{christfried.focke@northwestern.edu}
\affiliation{\Northwestern}

\author{Walter Giele}
\email[Electronic address:]{giele@fnal.gov}
\affiliation{\Fermilab}

\author{Xiaohui Liu}
\email[Electronic address:]{xhliu@umd.edu}
\affiliation{\Maryland}
\affiliation{\Peking}

\author{Frank Petriello}     
\email[Electronic address:]{f-petriello@northwestern.edu}
\affiliation{\Argonne}
\affiliation{\Northwestern}

\date{\today}
\begin{abstract}
We use the recently proposed jettiness-subtraction scheme to provide the complete calculation of Higgs boson production in association with a jet in hadronic collisions through next-to-next-to-leading order in perturbative QCD.  This method exploits the observation that the $N$-jettiness event-shape variable completely describes the singularity structure of QCD when final-state colored particles are present.  Our results are in agreement with a recent computation of the $gg$ and $qg$ partonic initial states based on sector-improved residue subtraction.  We present phenomenological results for both fiducial cross sections and distributions at the LHC.

\end{abstract}

\maketitle

\section{Introduction} \label{sec:intro}

The continued investigation of the Higgs boson will be a top priority of the high energy physics community 
during the coming years.  The comparison of Standard Model (SM) predictions with data from Run I of the 
Large Hadron Collider (LHC) was limited by the statistical precision of the experimental data.  This will 
no longer be the case during Run II, and systematic errors will dominate.  The largest systematic error currently 
hindering our understanding of Higgs properties is the theoretical understanding of the SM prediction.  This is the case 
for the well-measured di-boson modes~\cite{Higgserrors} which dominate the overall signal-strength determination.  
The theoretical uncertainties must be reduced in order to sharpen our understanding of the mechanism of electroweak 
symmetry breaking in Nature.

Improvements in both the overall production rate of the Higgs boson and in the modeling of its kinematic distributions 
are needed to match the expected experimental precision of Run II.  The total cross section is currently known at 
next-to-next-to-leading order (NNLO) in the strong coupling constant~\cite{Harlander:2002wh,Anastasiou:2002yz,Ravindran:2003um}, 
and the resummation of enhanced threshold logarithms at higher orders is known~\cite{Catani:2003zt,Moch:2005ky,Ahrens:2008nc}.  The N$^3$LO corrections have recently become available~\cite{Anastasiou:2015ema}, as are phenomenological predictions incorporating much of our current knowledge of the inclusive cross section~\cite{Bonvini:2014jma}.  To improve the modeling of Higgs kinematics at the LHC, precision predictions for Higgs production in association with jets are needed.  In 
decay modes such as $H \to W^+W^- \to l^+\bar{\nu}l^-\nu$ in which a mass peak cannot be reconstructed, the need for theory is critical.  The resummation of jet-binning logarithms beyond NLO in the $WW$ final state has been shown to lead to a factor of two reduction in 
the theory uncertainty affecting this channel~\cite{Boughezal:2013oha}, indicating that higher-order corrections play a vital role in the analysis of Higgs properties in this channel.

Initial results for Higgs production in association with a jet at NNLO are available, including both partial results in pure Yang-Mills~\cite{Boughezal:2013uia,Chen:2014gva} and a recent computation of the complete $gg$ and $qg$ partonic scattering channels~\cite{Boughezal:2015dra}.  We present here a full computation of all partonic channels through NNLO for the Higgs plus jet process using the recently proposed jettiness subtraction scheme~\cite{Boughezal:2015dva}.  Our aim in this manuscript is two-fold; first, we wish to demonstrate the application of jettiness subtraction in a non-trivial example.  Higgs production in association with a jet provides such a test case.  Second, given the importance of the precision Higgs program at the LHC, we wish to confirm the recent computations of this process.   We find agreement with the results obtained by the calculation in Ref.~\cite{Boughezal:2015dra}.  Our result in addition contains the quark-initiated scattering processes at NNLO, which were not included in previous calculations~\cite{Boughezal:2015dra}.  We confirm that they have minimal phenomenological impact.

Our paper is organized as follows.  We begin by reviewing the recently introduced jettiness-subtraction formalism.  The checks of our calculation, both internal ones and those involving other known results in the literature, are discussed next.  We then present numerical results for fiducial cross sections and for several distributions of both the Higgs boson and the leading jet.  We conclude with a summary and a discussion of potential future extensions of the work presented here.

\section{Review of Jettiness Subtraction} \label{sec:review}

We review here the salient features of the jettiness-subtraction scheme for NNLO calculations, which was recently 
introduced in the context of the NNLO computation of $W$-boson production in association with a jet~\cite{Boughezal:2015dva}.
We begin by defining the $N$-jettiness event shape variable $\taun$, first introduced in Ref.~\cite{Stewart:2010tn}:
\begin{equation}
{\cal T}_N = \sum_k \text{min}_i \left\{ \frac{2 p_i \cdot q_k}{Q_i}\right\}.
\end{equation}
The subscript $N$ denotes the number of jets in the final state.  For the $H$+jet process considered here, we have $N=1$.  Values of ${\cal T}_1$ near zero indicate a final state containing a single narrow energy deposition, while larger values denote a final state containing two or more well-separated energy depositions.  The first of these configurations will eventually reconstruct to one jet after imposing a jet algorithm, while the second will reconstruct to two or more jets.  The $p_i$ are light-like vectors for each of the initial beams and final-state jets in the problem, while the $q_k$ denote the four-momenta of any final-state radiation.  The $Q_i$ characterize the hardness of the beam-jets and final-state jets.  We set $Q_i = 2 E_i$, twice the energy of each jet.  

The cross section for $\taun$ less than some value $\tauncut$ can be expressed in the form~\cite{Stewart:2010pd,Stewart:2009yx}
\begin{equation} \label{eq:fact}
 \sigma(\taun < \tauncut)=  \int H \otimes B \otimes B \otimes S \otimes   \left[ \prod_n^{N} J_n \right] +\cdots .
\end{equation}
The function $H$ contains the virtual corrections to the process.  The beam function $B$ encodes the effect of radiation collinear to one of the two initial beams.  It can be written as a perturbative matching coefficient convoluted with a parton distribution function.  $S$ describes the soft radiation, while $J_n$ contains the radiation collinear to a final-state jet.  The ellipsis denotes power-suppressed terms which become negligible for $\taun \ll Q_i$. Each of these functions obeys a renormalization-group equation that allows logarithms of $\taun$ to be resummed.  If this expression is instead expanded to fixed-order in the strong coupling constant, it reproduces the cross section for low $\taun$. The derivation of this factorization theorem in the small-${\cal T}_N$ limit relies upon the machinery of Soft-Collinear Effective Theory~\cite{Bauer:2000ew}.

The basic idea behind jettiness subtraction is that $\taun$ fully captures the singularity structure of QCD amplitudes with final-state partons.  This allows us to calculate the NNLO corrections to processes such as Higgs plus jet in the following way.  We divide the phase space according to whether $\taun$ is greater than or less than $\tauncut$.  For $\taun > \tauncut$ there are at least two hard partons in the final state, since all singularities are controlled by jettiness.  This region of phase space can therefore be obtained from a NLO calculation of Higgs production in association with two jets.  Below $\tauncut$, the cross section is given by the factorization theorem of Eq.~(\ref{eq:fact}) expanded to second order in the strong coupling constant.  As long as $\tauncut$ is smaller than any other kinematic invariant in the problem, power corrections below the cutoff are unimportant.

All ingredients of Eq.~(\ref{eq:fact}) are known to the appropriate order to describe the low $\taun$ region through second order in the strong coupling constant.  The two-loop virtual corrections are known for Higgs plus jet~\cite{Gehrmann:2011aa}.
The beam functions are known through NNLO~\cite{Gaunt:2014xga,Gaunt:2014cfa}, as are the jet functions~\cite{Becher:2006qw,Becher:2010pd} and soft function~\cite{Boughezal:2015eha}.  It is therefore possible to combine this information to provide the full NNLO calculation of Higgs production in association with a jet.

\section{Validation of the Calculation} \label{sec:checks}

We describe here the various checks we have performed on our calculation.  For the phase space region above $\tauncut$ we need an NLO calculation of $H$+2-jets.  We use MCFM~\cite{Campbell:2010ff,Campbell:2015qma} for this purpose.  Below $\tauncut$, we require several separate terms.  We have checked our implementation of the two-loop virtual corrections against those contained in PeTeR~\cite{Becher:2014tsa}.  Our comparison of the soft function against known results in the literature is detailed in Ref.~\cite{Boughezal:2015eha}.

\begin{figure}[h]
\centering
\includegraphics[width=3.4in]{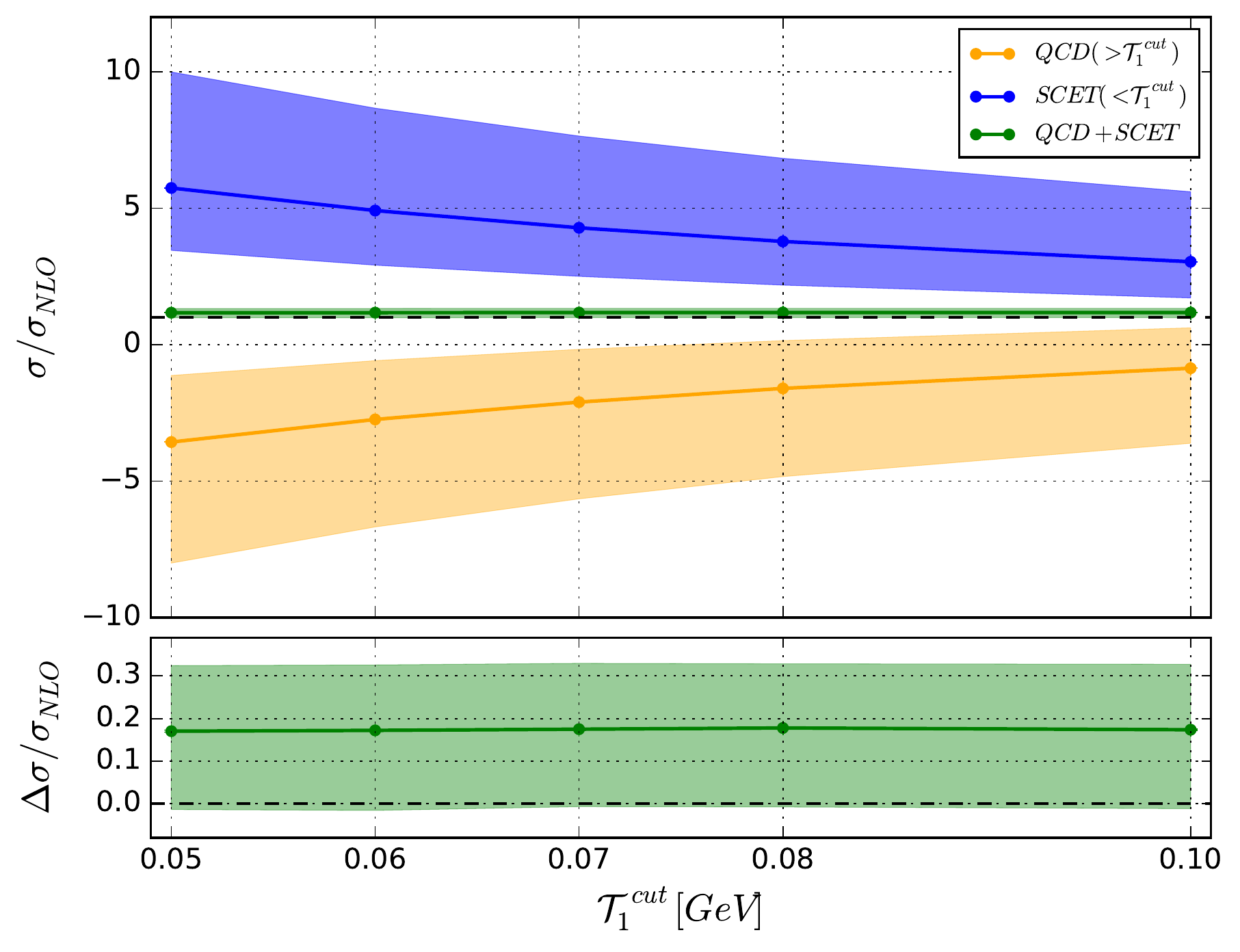}
\caption{The separate cross sections for the regions $\tauone>\tauonecut$ and $\tauone<\tauonecut$, together with their sum, as a function of $\tauonecut$, normalized to the NLO cross section.  The lower panel shows the relative correction also with respect to the NLO cross section.  The solid lines denote the result for the central scale choice $\mu=m_H$, while the bands show the result as the scale as varied in the range $m_H/2 \leq \mu \leq 2 m_H$.  The black dashed lines denote $\sigma/\sigma_{NLO}=1$ (upper panel) and 
$ \Delta \sigma/\sigma_{NLO}=0$ (lower panel). } \label{fig:taucheck}
\end{figure}

An important check of our formalism is the independence of our final result from $\tauonecut$, after both phase-space regions have been summed.  This requires a consistent implementation of the jet, beam, and soft functions below the cut, and the NLO result for $H$+2-jets above the cut.  This also allows us to define the appropriate range of $\tauonecut$ for which power corrections are negligible.  We show in Fig.~\ref{fig:taucheck} the NNLO correction to the cross section as a function of $\tauonecut$.  We also show separately the contributions above and below $\tauonecut$.  The cross sections from the separate regions are each a factor of several larger than their sum, and vary by more than a factor of two over the $\tauonecut$ range studied.  The sum is extremely stable over the entire range studied. 

\section{Numerical Results} \label{sec:numerics}

We now present numerical results for Higgs production in association with a jet.  We focus on 8 TeV proton-proton collisions in this paper.  Jets are reconstructed using the anti-$k_T$ algorithm~\cite{Cacciari:2008gp} with $R=0.5$.  We show results using the NNPDF~\cite{Ball:2012cx} parton distribution functions.  We use the perturbative order of the PDFs that is consistent with the partonic cross section under consideration: LO PDFs with LO partonic cross sections, NLO PDFs with NLO partonic cross sections, and NNLO PDFs with NNLO partonic cross sections.  We set the renormalization and factorization scales equal to the mass of the Higgs boson, $\mu_R = \mu_F = m_H$.  To estimate the residual theoretical error, we vary these scales simultaneously around this central value by a factor of two. We set the mass of the Higgs boson as $m_H = 125$ GeV.  

We begin by studying the fiducial cross section for Higgs+jet production, which we define by imposing the following cuts on the final state jet: $p_T^{jet}> 30$ GeV, $|Y^{jet}|<2.5$.  Our results are shown in Table~\ref{tab:fiducial}.  We can compare these results against the calculation of Ref.~\cite{Boughezal:2015dra}, which is based on the technique of sector-improved residue subtraction~\cite{Czakon:2010td,Boughezal:2011jf}. The result quoted in Ref.~\cite{Boughezal:2015dra} does not include a cut on $|Y^{jet}|$, and also does not include the quark-initiated partonic channels $qq$, $\bar{q}q$, and $\bar{q}\bar{q}$.  Incorporating the $|Y^{jet}|$ cut in the sector-improved subtraction calculation~\cite{privcomm} and removing the quark-initiated channels from our result, we find agreement within numerical errors.  We note that the quark-initiated partonic channels reduce the NNLO result by approximately 1.5\% within the fiducial region studied here, indicating that they have small phenomenological impact.

\begin{table}[h]
\begin{tabular}{|l|l|}
\hline
\multicolumn{2}{|c|}{$p_T^{jet}> 30$ GeV, $|Y^{jet}|<2.5$}\\ \hline\hline
Leading order: & $3.1^{+1.3}_{-0.9}$ pb \\ \hline
Next-to-leading order: & $4.8^{+1.1}_{-0.9}$ pb \\ \hline
Next-to-next-to-leading order: & $5.5^{+0.3}_{-0.4}$ pb \\ \hline
\end{tabular} \caption{Fiducial cross sections, defined by $p_T^{jet}> 30$ GeV, $|Y^{jet}|<2.5$, using NNPDF PDFs at each order of perturbation theory.   The central scale choice is $\mu=m_H$.  Results for $\mu = m_H/2$ and $\mu = 2m_H$ are shown as superscripts and subscripts, respectively. } \label{tab:fiducial}
\end{table}

\begin{figure}[h]
\centering
\includegraphics[width=3.4in]{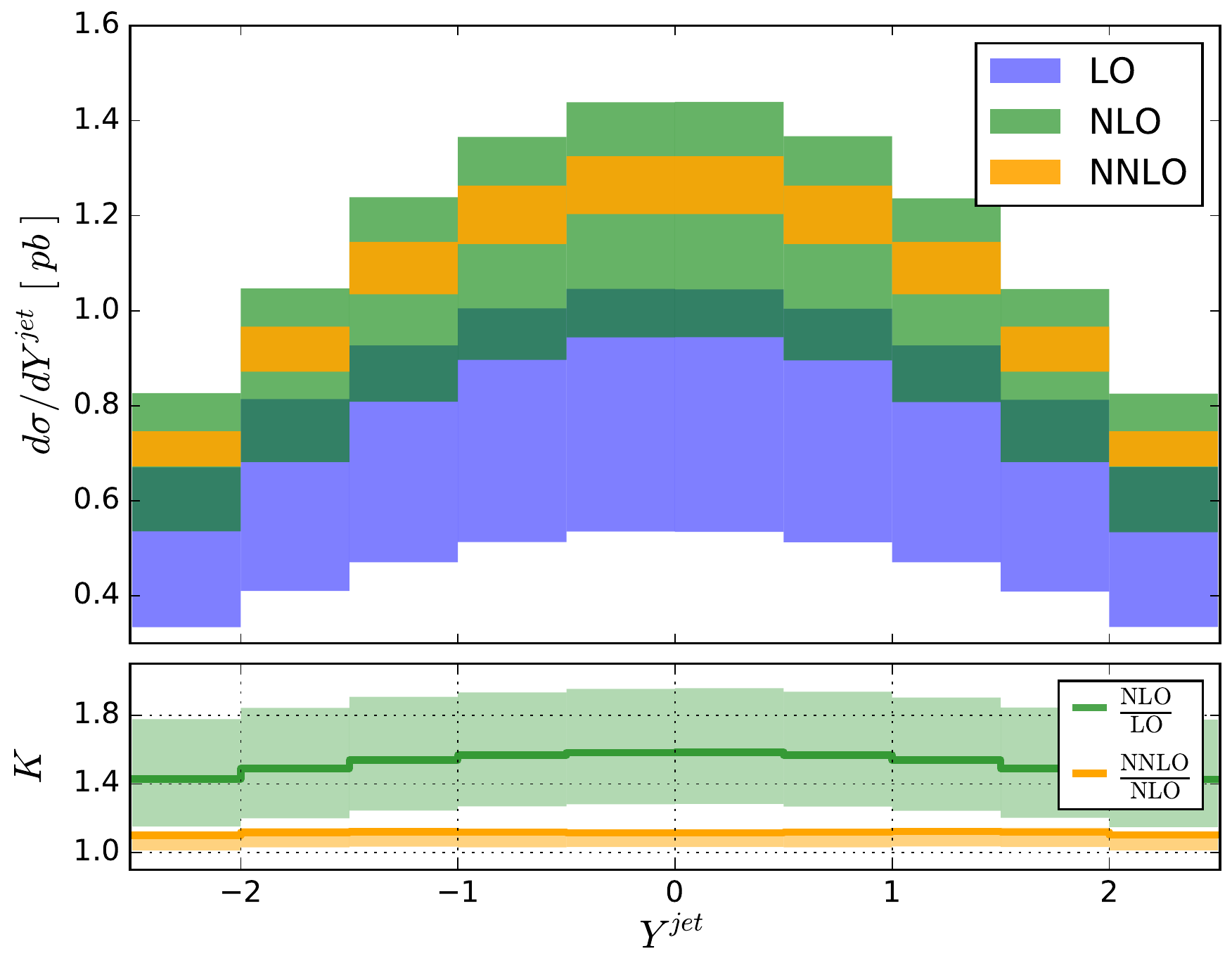}
\caption{The rapidity of the leading jet at LO, NLO, and NNLO in the strong coupling constant. The lower inset shows the ratios of NLO over LO cross sections, and NNLO over NLO cross sections. Both shaded regions in the upper panel and the lower inset indicate the scale-variation errors.} \label{fig:Yjet}
\end{figure}

\begin{figure}[h]
\centering
\includegraphics[width=3.4in]{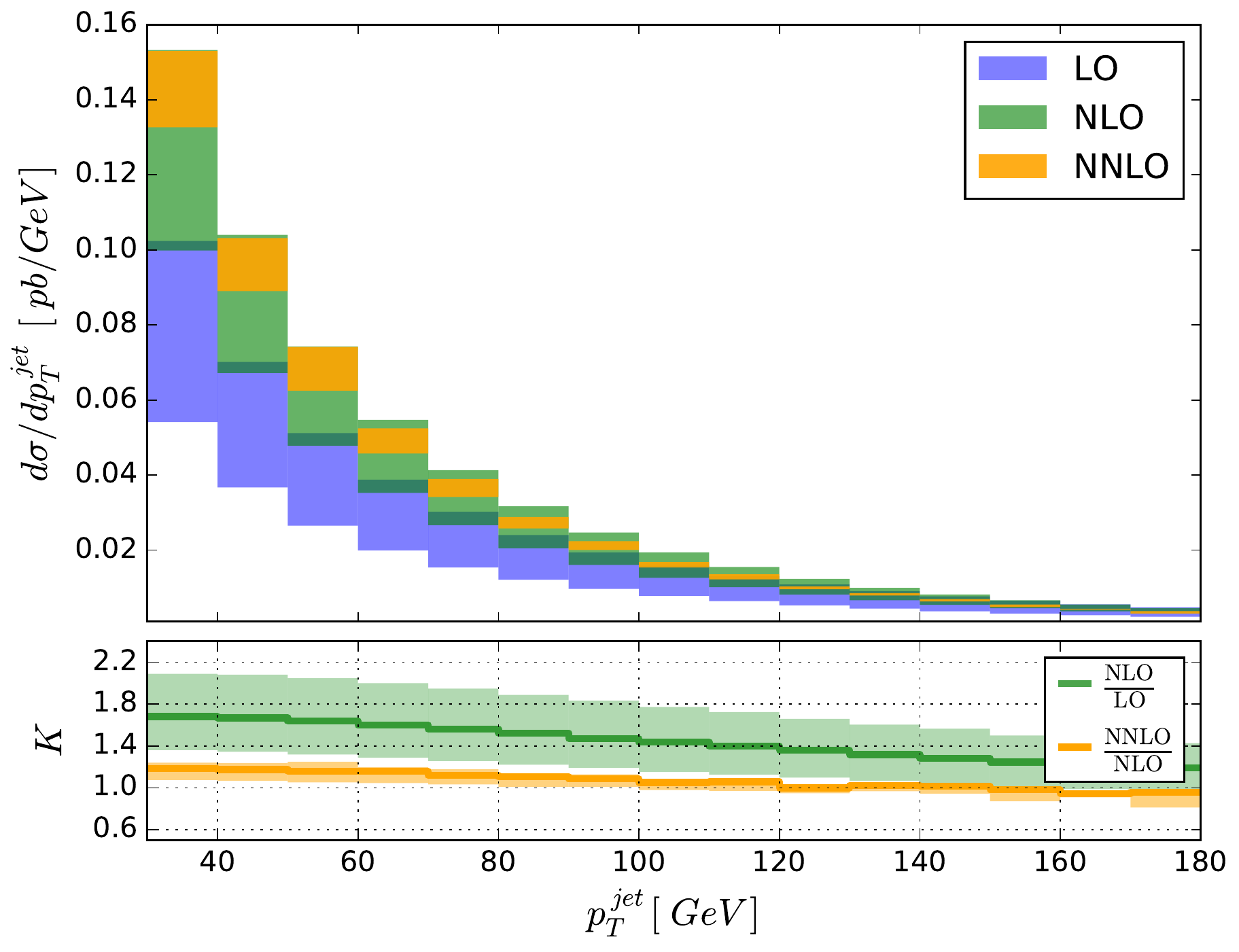}
\caption{The transverse momentum of the leading jet at LO, NLO, and NNLO in the strong coupling constant. The lower inset shows the ratios of NLO over LO cross sections, and NNLO over NLO cross sections.  Both shaded regions in the upper panel and the lower inset indicate the scale-variation errors. } \label{fig:pTjet}
\end{figure}

We next show several distributions in Higgs plus jet production.  We show in Fig.~\ref{fig:Yjet} the rapidity distribution of the leading jet at each order in perturbation theory, as well as the $K$-factors (defined as ratios of higher-order cross section over the lower-order ones) in the lower inset.  The NLO corrections exhibit a slight shape dependence, with the corrections approximately 10-20\% larger in the central region than near $|Y^{jet}|=2.5$.  The NNLO corrections are flatter as a function of rapidity, and the NNLO distribution is entirely contained within the NLO scale variation band.  In Fig.~\ref{fig:pTjet} we show the transverse momentum distribution of the leading jet. There is a shape dependence to the corrections, with the $K$-factor decreasing as $p_T^{jet}$ is increased. This trend is visible when going from LO to NLO in perturbation theory, and also when going from NLO to NNLO.  We note that the NNLO result is entirely contained within the NLO scale-variation band.  The shape dependence and magnitude of the NNLO corrections for the $p_T^{jet}$ distribution are in agreement with the results of Ref.~\cite{Boughezal:2015dra}.

\begin{figure}[h]
\centering
\includegraphics[width=3.4in]{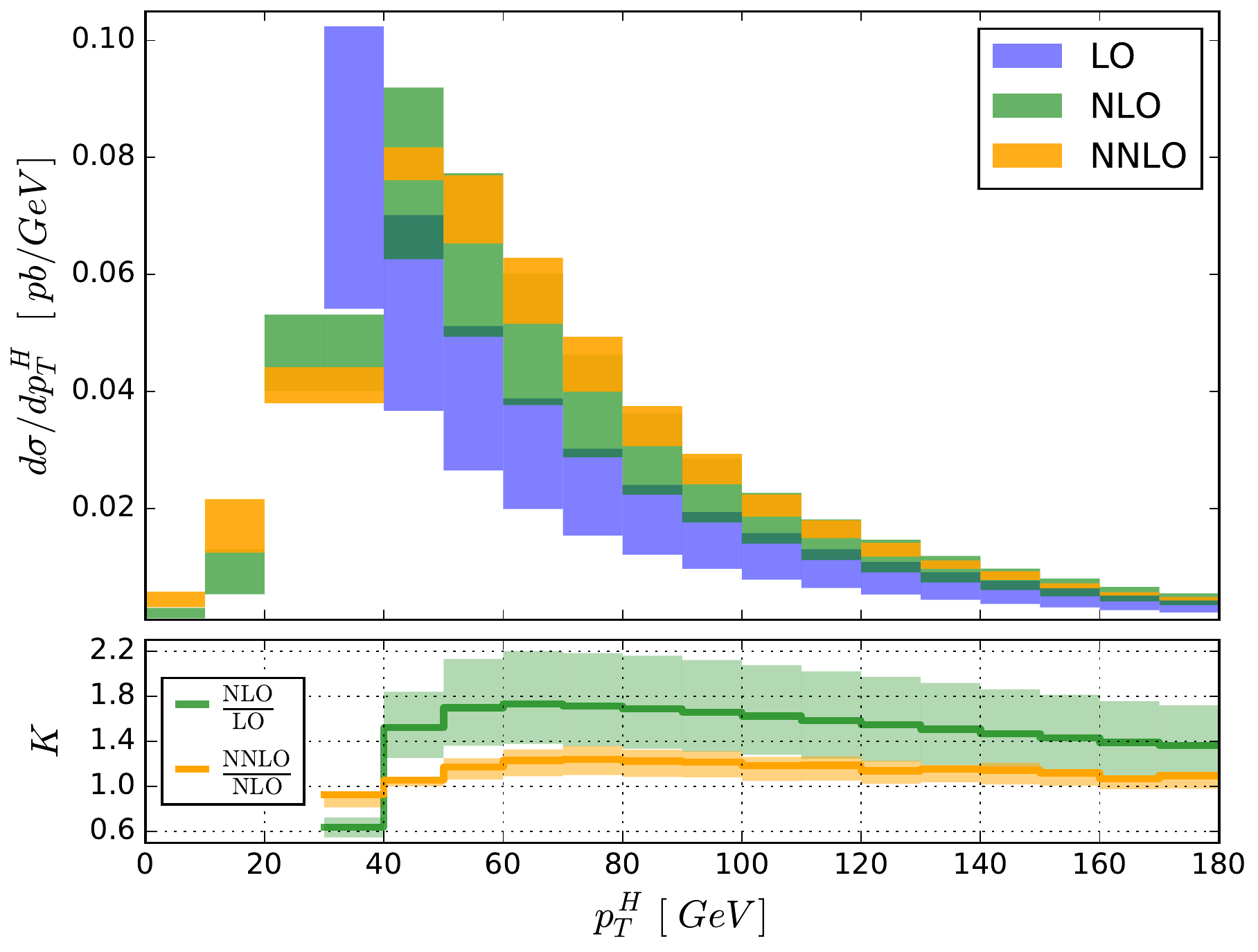}
\caption{The transverse momentum of the Higgs boson at LO, NLO, and NNLO in the strong coupling constant. The lower inset shows the ratios of NLO over LO cross sections, and NNLO over NLO cross sections.  Both shaded regions in the upper panel and the lower inset indicate the scale-variation errors.} \label{fig:pThiggs}
\end{figure}

Finally, we show in Fig.~\ref{fig:pThiggs} the transverse momentum of the Higgs boson in the $H$+jet process. The NLO corrections range from 40\% to 120\% near $p_{T}^H=60$ GeV, depending on the scale choice.  The magnitude of this correction decreases as the transverse momentum of the Higgs increases.  The NNLO corrections are more mild, reaching only 20\% at most for the central scale choice $\mu=m_H$.  They also decrease slightly as the transverse momentum of the Higgs increases.  The shape dependence and magnitude of the NNLO corrections for the $p_T^{H}$ distribution are in agreement with the results of Ref.~\cite{Boughezal:2015dra}. 
We note that we have combined the two bins closest to the boundary $p_T^H=30$ GeV to avoid the well-known Sudakov shoulder effect~\cite{Catani:1997xc}.

\section{Conclusions} \label{sec:conc}

We have presented in this manuscript a complete calculation of Higgs production in association with a jet through NNLO in perturbative QCD.  Our computation uses the recently proposed method of jettiness subtraction, a general technique for obtaining higher-order corrections to processes containing final-state jets.  We confirm and extend a recent calculation of the dominant $gg$ and $qg$ partonic channels through NNLO~\cite{Boughezal:2013uia}, and present additional phenomenological results for 8 TeV LHC collisions.   We also present several distributions for the Higgs and the leading jet that can be measured with LHC data.  Our results indicate that the perturbative series is under good control after the inclusion of the NNLO corrections. We look forward to the comparison of our
theoretical prediction with the upcoming data from Run II of the LHC.

\section{Acknowledgements}

R.~B. is supported by the DOE contract DE-AC02-06CH11357.  C.~F. is supported by the DOE grant DE-FG02-91ER40684.  W.~G. is supported by the DOE contract DE-AC02-07CH11359.  X.~L. is supported by the DOE.  F.~P. is supported by the DOE grants DE-FG02-91ER40684 and DE-AC02-06CH11357.  This research used resources of the National Energy Research Scientific Computing Center, a DOE Office of Science User Facility supported by the Office of Science of the U.S. Department of Energy under Contract No. DE-AC02-05CH11231.

\end{document}